\documentclass[aps,groupedaddress,superscriptaddress, pra,twocolumn,a4paper,floatfix,10pt,longbibliography]{revtex4-1}

\usepackage{amsmath}
\usepackage{enumitem}
\usepackage{amssymb}
\usepackage{amsfonts}
\usepackage{braket}
\usepackage{dsfont}
\usepackage{MnSymbol} 

\usepackage{color}
\usepackage[dvipsnames]{xcolor}
\usepackage{dirtytalk}
\usepackage{amsmath}
\usepackage{tikz}

\DeclareMathOperator{\tr}{tr}

\usepackage{graphicx, import}
\usepackage{soul}
\usepackage[utf8]{inputenc}
\usepackage[export]{adjustbox}
\usepackage[caption=false]{subfig}
\usepackage{parskip} 
\usepackage{hyperref}

\usepackage[textsize=tiny]{todonotes}

\usepackage{xspace}

\newcommand{\overbar}[1]{\mkern10.0mu\overline{\mkern-10.0mu#1\mkern-10.0mu}\mkern 10.0mu}

\begin{document}
\title{Quantum Self-Supervised Learning}

\author{B. Jaderberg}
\thanks{These authors contributed equally to this work}
\affiliation{Clarendon Laboratory, University of Oxford, Parks Road, Oxford OX1 3PU, United Kingdom}

\author{L. W. Anderson}
\thanks{These authors contributed equally to this work}
\affiliation{Clarendon Laboratory, University of Oxford, Parks Road, Oxford OX1 3PU, United Kingdom}

\author{W. Xie}
\affiliation{Visual Geometry Group, Department of Engineering Science, University of Oxford}

\author{S. Albanie}
\affiliation{Department of Engineering, University of Cambridge}

\author{M. Kiffner}
\affiliation{Clarendon Laboratory, University of Oxford, Parks Road, Oxford OX1 3PU, United Kingdom}
\affiliation{Centre for Quantum Technologies, National University of Singapore, 3 Science Drive 2, Singapore 117543}

\author{D. Jaksch}
\affiliation{Clarendon Laboratory, University of Oxford, Parks Road, Oxford OX1 3PU, United Kingdom}
\affiliation{Centre for Quantum Technologies, National University of Singapore, 3 Science Drive 2, Singapore 117543}
\affiliation{Institut für Laserphysik, Universität Hamburg, 22761 Hamburg, Germany}

\date{\today}

\begin{abstract}

The resurgence of self-supervised learning, whereby a deep learning model generates its own supervisory signal from the data, promises a scalable way to tackle the dramatically increasing size of real-world data sets without human annotation. However, the staggering computational complexity of these methods is such that for state-of-the-art performance, classical hardware requirements represent a significant bottleneck to further progress. Here we take the first steps to understanding whether quantum neural networks could meet the demand for more powerful architectures and test its effectiveness in proof-of-principle hybrid experiments. Interestingly, we observe a numerical advantage for the learning of visual representations using small-scale quantum neural networks over equivalently structured classical networks, even when the quantum circuits are sampled with only 100 shots. Furthermore, we apply our best quantum model to classify unseen images on the \textit{ibmq\_paris} quantum computer and find that current noisy devices can already achieve equal accuracy to the equivalent classical model on downstream tasks.

\end{abstract}

\maketitle

\section{Introduction} \label{sec:introduction}

In the past decade, machine learning has revolutionised scientific analysis, yielding breakthrough results in protein folding~\cite{senior2020improved}, black hole imaging~\cite{Akiyama2019FirstME} and heart disease treatment~\cite{theodoris2021network}. At the forefront of this progress is deep learning~\cite{lecun2015deep}, characterised by the successive application of artificial neural network layers~\cite{mcculloch1943logical, krogh2008artificial}. Notably, its use in computer vision has seen the top-1 accuracy on benchmark datasets such as ImageNet soar from $52\%$~\cite{lin2011large} to over $90\%$~\cite{pham2021meta}, fuelled by shifts in the underlying techniques used~\cite{lowe1999object, krizhevsky2012imagenet}. However, what has remained consistent in these top performing models is the use of labelled data to supervise the representation learning process. Whilst effective, the reliance on large quantities of human-provided annotations presents a significant challenge as to whether such approaches will scale into the future. Crucially, modern datasets such as the billions of images uploaded to social media are both vast and unbounded in their subject, quickly making the task of labelling unfeasible.

This has reignited interest in an alternative approach, termed \textit{self-supervised learning}~\cite{de1994learning}, which seeks instead to exploit structure in the data itself as a learning signal. Rather than predict human annotations, a model is trained to perform a \textit{proxy task}, that makes use of attributes of the data that can be inferred without labelling. Furthermore, the proxy task should encourage the model to learn representations that capture useful factors of variation in the visual input, such that solving it ultimately correlates with solving tasks of interest after training. Recent progress in the self-supervised learning of visual data has been driven by the success of contrastive learning~\cite{wu2018improving, wu2018unsupervised, henaff2020data, oord2018representation, he2020momentum}, in which the proxy task is differentiating augmented instances of the same image from all other images. Provided the correct choice of augmentations, this produces a model which is invariant to transformations that do not change the semantic meaning of the image, allowing the learning of recognisable features and patterns in unlabelled datasets. 

With these techniques, contrastive learning is able to learn visual representations with comparable quality to supervised learning~\cite{chen2020simple,he2020momentum}, without the bottleneck of labelling. However, it is a fundamentally more difficult task than its supervised counterpart~\cite{wang2021solving}, and capturing complex correlations between augmented views requires more training data, more training time and larger network capacity~\cite{henaff2020data, oord2018representation}. Therefore, it is important to consider whether emerging technologies can contribute to the growing requirement for more powerful neural networks~\cite{he2021masked}.

Variational quantum algorithms (VQAs)~\cite{cerezo2021variational}, a near term application of quantum computing, are one such new paradigm. While VQAs have been used to solve many types of optimisation problems~\cite{peruzzo2014variational, google2020hartree, farhi2014quantum, zhou2020quantum, ma2020quantum}, it is their application to supervised learning~\cite{grant2018hierarchical,vojtech2019supervised,schuld2020circuit}, unsupervised learning~\cite{otterbach2017unsupervised}, generative models~\cite{benedetti2019generative,zoufal2019quantum} and reinforcement learning~\cite{chen2020variational,lockwood2020reinforcement, saggio2021experimental} which has led to them being referred to as quantum neural networks (QNNs)~\cite{mitari2018quantum, beer2020training, benedetti2019parameterized}. In theory, the power of these models comes from their access to an exponentially large feature space~\cite{vojtech2019supervised} and ability to represent complex high-dimensional distributions, as formalised by the effective dimension~\cite{abbas2021power}. Importantly, early evidence suggests that quantum models can achieve an advantage over their classical counterparts, yet these works focus only on the supervised learning of either artificial data~\cite{beer2020training, huang2021power} or simple historical datasets~\cite{abbas2021power}. For example, whilst widely used to study QNNs~\cite{bausch2020recurrent, skolik2021layerwise}, classical supervised learning of MNIST can already achieve 99.3\% top-1 accuracy with a two-layer 784-800 width multi-layer perceptron (MLP)~\cite{simard2003best}. Thus, it is highly unlikely that this problem would practically benefit from a quantum model with access to a $>2^{50}$ dimensional feature space and careful consideration should be made about whether supervised learning is the best setting to try to achieve quantum advantage. By comparison, self-supervised learning of ImageNet with the widely-used ResNet 50 architecture~\cite{he2016deep} (with maximum channel width 2048) achieves only 76.5\% top-1 accuracy~\cite{chen2020simple}. The necessity for large capacity models means that self-supervised learning may be a better setting in which to seek useful quantum advantage through quantum neural networks~\cite{abbas2021power}.

In this work, we construct a contrastive learning architecture in which classical and quantum neural networks are trained together. By randomly augmenting each image in the dataset, our hybrid network learns visual representations which groups different views of the same image together in both classical and Hilbert space. Afterwards, we test the quality of the representations by using them to train a linear classifier, which then makes predictions on an unseen test set. We find that our hybrid encoder, constrained in both size and training time by quantum simulation overheads, achieves an average test accuracy of $(46.51\pm1.37)\%$. In contrast, replacing the QNN with a classical neural network of equivalent width and depth results in a model which obtains $(43.49\pm1.31)\%$ accuracy. Thus, our results provide the first indication that a quantum model may better capture the complex correlations required for self-supervised learning.

We then apply the best performing quantum model to classify test images on a real quantum computer. Notably, the accuracy achieved using the \textit{ibmq\_paris}~\cite{ibmq_systems} device equals the best performing classical model, despite significant device noise. This illustrates the capability of our algorithm for real-world applications using current devices, with flexibility to assign more of the encoding to QNNs as quantum hardware improves. While further research is required to demonstrate scalability, our scheme provides a strong foundation for quantum self-supervised learning. Excitingly, given that contrastive learning has also been successfully applied to non-visual data~\cite{oord2018representation, Mnih13, Grover16, lu2020self, jaeger2018mol2vec}, our work opens the possibility of using QNNs to learn large, unlabelled datasets across a range of disciplines.
	
\section{Method}\label{sec:method}

\subsection{Contrastive learning architecture}\label{subsec:architecture}

\begin{figure}
	\centering
	\includegraphics[width=0.85\linewidth]{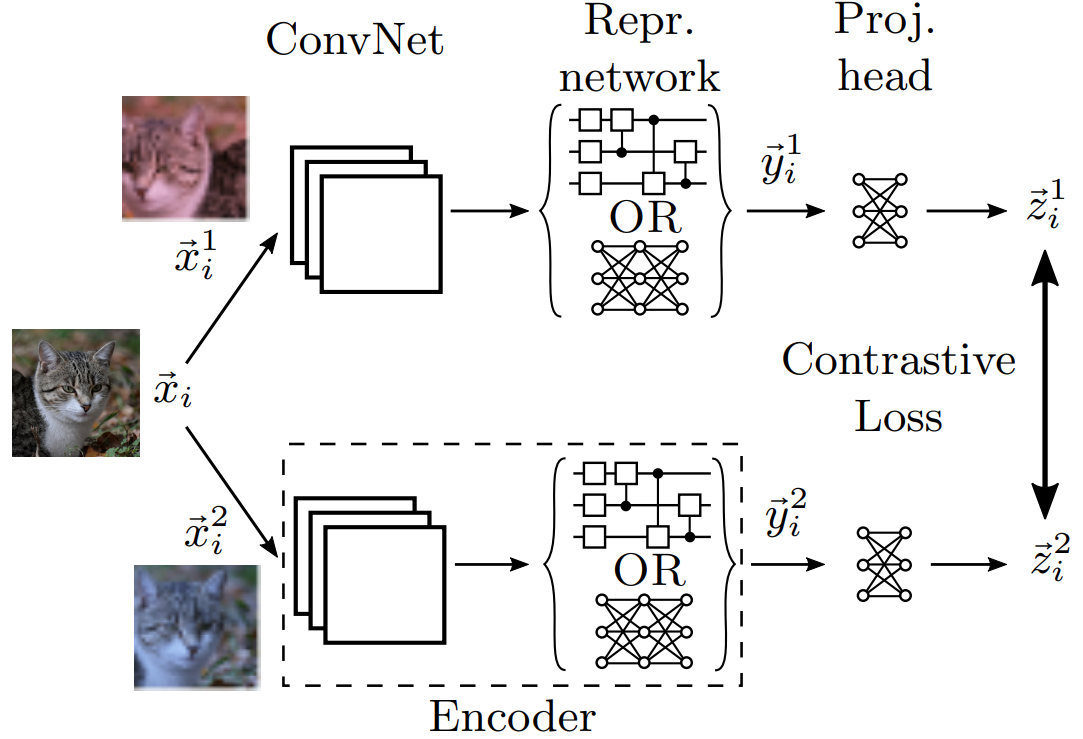}
	\caption{Schematic of the overall neural network architecture and contrastive training method. For each input image $\vec{x}_i$, a pair of random augmentations are generated and applied to form a positive pair $\vec{x}_i^{1}$, $\vec{x}_i^{2}$. These are transformed by the encoder network, consisting of classical convolutional layers and a quantum or classical representation network, into representation vectors $\vec{y}_i^{1}$, $\vec{y}_i^{2}$. The projection head subsequently maps the representations to the vectors $\vec{z}_i^{1}$, $\vec{z}_i^{2}$, such that contrastive loss can be applied without inducing loss of information on the encoder.}
	\label{fig:network_architecture}
\end{figure}

Given an unlabelled dataset, the objective of self-supervised learning is to find low dimensional encodings of the images which retain important higher level features. In this work, we train a model to do this by adapting the widely used SimCLR algorithm~\cite{chen2020simple}, the steps of which can be seen in Fig. \ref{fig:network_architecture}. Firstly for a given image, the data of which is contained within $\vec{x}_i$, we generate two augmentation functions. Each one randomly crops, rotates, blurs and colour distorts the picture, such that two augmented views $\vec{x}_i^{1}$, $\vec{x}_i^{2}$ of the same base image are produced. Importantly, these augmentations still allow for the underlying object to remain visually distinguishable. This enables us to assert that these two views contain a recognisable description of the same class, which we call a positive pair. 

Once this positive pair is generated, each view is passed through a set of neural networks. First, an encoder network is applied, which maps the high dimensional input data $\vec{x}_i^{1}$, $\vec{x}_i^{2}$ to low dimensional representations $\vec{y}_i^{1}$, $\vec{y}_i^{2}$. Then the output of the encoder network is passed to the projection head, a small multi-layer-perceptron (MLP)~\cite{du2013neural} consisting of two fully connected layers. This produces the final representations $\vec{z}_i^{1}$, $\vec{z}_i^{2}$.

Given a batch of $N$ images, the above process is repeated such that we are left with $2N$ representations corresponding to $2N$ augmented views. Looking at all possible pairings of these representations, we have not only positive pairs (e.g., $\vec{z}_i^{1}$, $\vec{z}_i^{2}$) but also negative pairs (e.g., $\vec{z}_i^{1}$, $\vec{z}_j^{1}$ where $i\neq j$), which we cannot definitely say contain the same class. For each training step, all of these possible pairs are used to calculate the normalised temperature-scaled cross entropy loss (NT-Xent)~\cite{sohn2016improved} (see Appendix \ref{app:contrasitve_loss}), which is minimised via stochastic gradient descent~\cite{robbins1951stochastic}. Intuitively, minimising this loss function can be understood as training the network to produce representations in which positive pairs are mapped close together and negative pairs far apart, as measured by their cosine similarity. This idea is a core concept in contrastive learning and many machine learning techniques~\cite{hadsell2006dimensionality}. Note that whilst it is possible to train the network by applying NT-Xent directly to the output of the encoder, the contrastive loss function is known to induce loss of information on the layer it is applied to~\cite{chen2020simple}. Therefore, the addition of the projection head ensures that the encoder remains sensitive to image characteristics (e.g., colour, orientation) that improves performance on downstream tasks. 

In order to incorporate QNNs, we modify the encoder to contain both classical and quantum layers working together. The first part of the encoder consists of a convolutional neural network, which in this work is the widely used ResNet-18. This produces a 512 length feature vector, which is already an initial encoding of the augmented image. However, we then extend the encoder with a second network, which we call the representation network as it acts directly on the representation space. This consists of either a multi-layer QNN of width $W$, or a classical fully connected MLP with equivalent width and depth. Ideally the representation network would have width $W=512$, so as to minimise loss of information. However, we instead look to work in a regime which is realisable on current quantum computers, and as such in this work we use $W=8$. This is achieved by following the convolutional network with a single classical layer that compresses the vector, a common technique used to link classical and quantum networks together~\cite{mari2020transfer, lloyd2020quantum}.

After the representation network is applied, the resultant encoding is passed onto the previously described projection head. To maintain the structure of the original SimCLR architecture, we limit the projection head to be no wider than the width of the QNN.

\subsection{Quantum representation network}

\begin{figure}
	\includegraphics[width=0.9\linewidth]{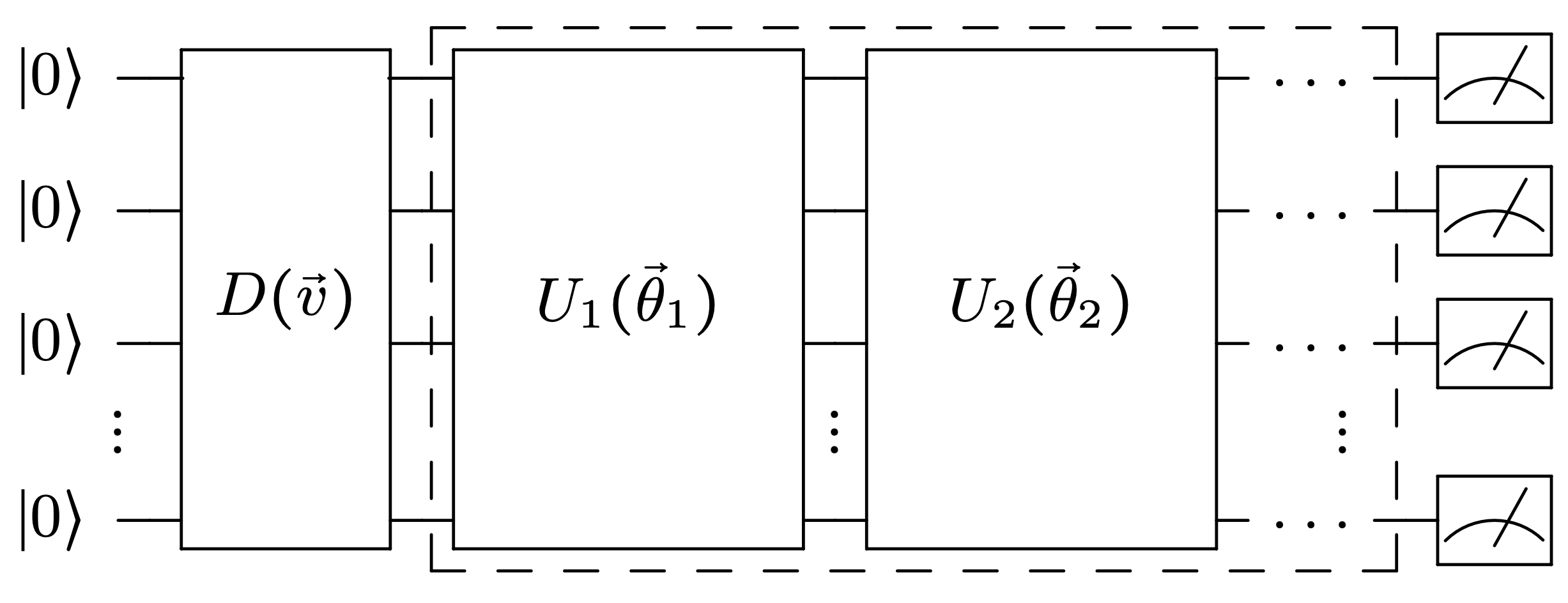}
	\caption{General structure of a QNN. An input vector $\vec{v}$ is encoded into the qubits by a data loading unitary $\hat{D}(\vec{v})$. The variational ansatz consists of layers $\set{\hat{U}_1(\vec{\theta}_1), \hat{U}_2(\vec{\theta}_2), \dots}$ and is parameterised by trainable parameters $\set{\vec{\theta}_1, \vec{\theta}_2, \dots}$. The output of the QNN is taken as the average of repeated measurements in the $\hat{\sigma}_{z}$ basis.}
	\label{fig:qnn_diagram}
\end{figure}

The quantum representation network follows the structure shown in Fig. \ref{fig:qnn_diagram}, beginning with a data loading unitary $\hat{D}(\vec{v})$. Whilst schemes exist to encode data into quantum circuits with exponential compression~\cite{le2011flexible,zhang2013neqr}, these require a prohibitively large number of logic gates compared to current hardware capabilities. By compressing the output of the ConvNet as described in section \ref{subsec:architecture}, we need only to solve the simpler issue of loading a vector $\vec{v}$ of length $W$ into equally as many qubits. This is achieved by applying a single qubit rotation $\hat{R}_x$ to each qubit in the register; $\hat{D}(\vec{v}) = \bigotimes_{i=1}^W \hat{R}_x(v^{k})$. Here, $v^k$ is the $k$th element of input vector $\vec{v}$ and is mapped to the range $[0, \pi]$ to prevent large values wrapping back around the Bloch sphere.

\begin{figure}
	\centering
	\includegraphics[width=0.8\linewidth]{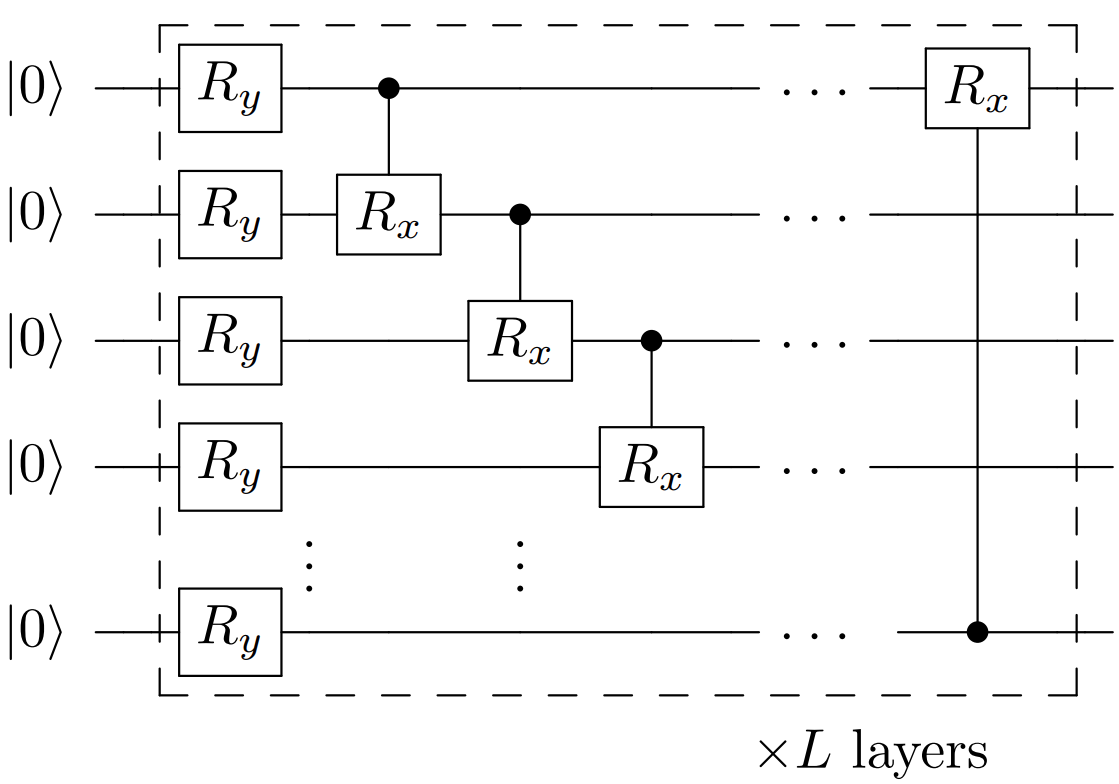}
	\caption{Variational ansatz used in this work. Each layer consists of a single qubit $\hat{R}_y$ rotation on each qubit, followed by controlled $\hat{R}_x$ rotations, connecting the qubits in a ring topology. Every rotation gate is parameterised by a different variational parameter.}
	\label{fig:sim_circ_14}
\end{figure}

Once the input data is loaded, we apply the learning component of our QNN, a parameterised quantum circuit ansatz. In applications where the ansatz is used to solve optimisation problems relating to a physical system (e.g., the simulation of molecules), the circuit structure and choice of logic gates can be inspired by the underlying Hamiltonian~\cite{grimsley2019adaptive}. However, without such symmetries to guide our choice, we use a variational ansatz based on recent theoretical findings in expressibility and entangling capability~\cite{sim2019expressibility}. The ansatz is shown in Fig. \ref{fig:sim_circ_14}, the structure of which is derived from circuit 14 of Ref.~\cite{sim2019expressibility} and was chosen due to its performance in both these metrics.

After the application of several ansatz layers, the network is finished by measuring each qubit to obtain an expectation value in the $\hat{\sigma}_{z}$ basis. When evaluated on a real quantum computer or sampling-based simulator, the expectation value is constructed by averaging the sampled eigenvalues over a finite number of shots. If evaluated on a statevector simulator, the expectation value is calculated exactly.

The gradients of the QNN output with respect to the trainable parameters and the input parameters are calculated using the parameter shift rule~\cite{mitari2018quantum, schuld2019evaluating}, which we describe here. Consider an observable $\hat{O}$ measured on the state
\begin{equation}
    \ket{\psi(\vec{\theta})} = \prod_i \hat{U}_i(\theta_i)\hat{V}_i \ket{0},
\end{equation}
resulting from the application of $M$ parameterised gates $\hat{U}_1,\hat{U}_2,\dots, \hat{U}_M$ and $M$ fixed gates $\hat{V}_1,\hat{V}_2,\dots, \hat{V}_M$, where gates $\hat{U_i} = e^{i\theta_i\hat{P}_i/2}$ are generated by operators $\hat{P}_i\in\set{\mathds{1},\hat{\sigma}_x, \hat{\sigma}_y, \hat{\sigma}_z }^{\otimes n}$ that are tensor products of the Pauli operators. According to the parameter shift rule, the gradient of the expectation value $f = \bra{\psi(\vec{\theta})} \hat{O} \ket{\psi(\vec{\theta})}$ with respect to parameter $\theta_i$ is given by
\begin{equation}\label{eqn:param_shift}
    \frac{\partial f(\vec{\theta})}{\partial \theta_i} = \frac{1}{2}\left[f\left(\theta_i + \frac{\pi}{2}\right) - f\left(\theta_i -\frac{\pi}{2}\right)\right]. 
\end{equation}

For each parameterised gate within the circuit, including both the variational ansatz and data loading unitary, an unbiased estimator for the gradient is calculated by measuring the QNN with the two shifted parameter values given in Eq. (\ref{eqn:param_shift}).

Once the QNN gradients have been calculated, we combine them with gradients of the classical components to obtain gradients of the loss function with respect to all trainable quantum and classical parameters via backpropagation~\cite{lecun1998gradient}. In this way, the QNN is trained simultaneously with the classical networks, and the quality of the gradients produced on quantum hardware play a crucial role in the training ability of the whole network.

\section{Results}\label{sec:results}

\subsection{Training}\label{subsec:training}
To examine whether the proposed architecture can successfully train, we apply it to the CIFAR-10 dataset~\cite{krizhevsky2009learning}. In this preliminary experiment we restrict the dataset to the first two classes, leaving 10,000 32$\times$32 colour images containing either an aeroplane or automobile. We also train this initial model without a projection head, since it is not being used for classification later. The quantum representation network here is a simulated two-layer QNN and is trained together with the classical components from scratch by integrating~\cite{jaderberg2021quantum} the Qiskit~\cite{Qiskit} and PyTorch~\cite{pytorch} frameworks. The full list of training hyperparameters can be found in Appendix \ref{app:hyperparameters}.

\begin{figure}
	\centering
	\includegraphics{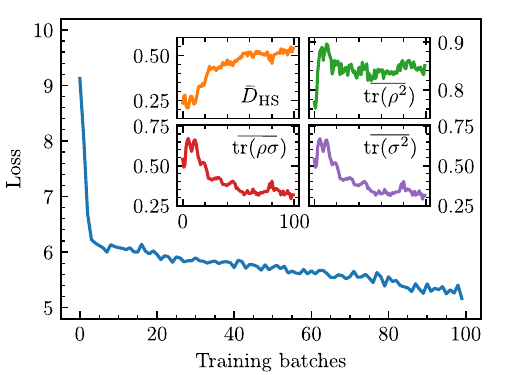}
	\caption{Contrastive learning with a quantum representation network. After each batch of 256 images the loss function (main graph) is recorded, alongside the average Hilbert Schmidt distance between positive and negative pairs $\bar{D}_{\text{HS}}$, the average positive pair clustering $\overbar{\text{tr}(\rho^2)}$, the average clustering of all negative pairs $\overbar{\text{tr}(\sigma^2)}$ and the ensemble inter-cluster overlap $\overbar{\text{tr}(\rho\sigma)}$ (insets).}
	\label{fig:training}
\end{figure}

Fig. \ref{fig:training} shows the results of several key metrics after training for 100 batches. Firstly, we record the loss after each batch, the minimisation of which represents the ability to produce representations in the classical $W$ dimensional space whereby positive pairs have high similarity. Our results show that the loss decreases from 9.13 to 5.16 over the course of training, indicating that our model is able to learn. Importantly, since the quantum and classical parameters are trained together, this shows that information is successfully passed both forwards and backwards between these different network paradigms.

Secondly, we log the Hilbert-Schmidt distance ($D_{\text{HS}}$), a metric that has been applied in quantum machine learning previously to study data embedding in Hilbert space~\cite{lloyd2020quantum}. Here, we use it to track the separation between our pseudo classes in the $2^{W}$ dimensional quantum state space while optimising the classical loss function. For a given positive pair $\vec{x}_{i}^{1}, \vec{x}_{i}^{2}$, we calculate the statistical ensembles 

\begin{subequations}
\begin{align}\label{rho}
\rho_i &= \frac{1}{2}\left(\ket{\psi_i^1}\bra{\psi_i^1} + \ket{\psi_i^2}\bra{\psi_i^2}\right),\\
\label{sigma}
\sigma_i &= \frac{1}{2N-2}\sum_{j\neq i} \left(\ket{\psi_j^1}\bra{\psi_j^1} + \ket{\psi_j^2}\bra{\psi_j^2}\right),
\end{align}
\end{subequations}

where $\ket{\psi_i^\alpha}$ is the statevector produced by the hybrid encoder given augmented view $\vec{x}_{i}^{\alpha}$. The Hilbert-Schmidt distance is then given by 

\begin{equation} \label{eqn:hilbert-schmidt}
D_{\text{HS},i} = §((\rho_i - \sigma_i)^2).
\end{equation}

We repeat this for each positive pair in the batch and record the mean, $\bar{D}_{\text{HS}} = \frac{1}{N}\sum_iD_{\text{HS},i}$. Focusing on the inset of Fig. \ref{fig:training}, we see in the upper-left panel that $\bar{D}_{\text{HS}}$ increases consistently across the range of training, indicating that the QNN successfully learns to separate positive and negative pairs in Hilbert space. Expanding out the quadratic in Eq. (\ref{eqn:hilbert-schmidt}), we can break down the metric into the so-called purity terms $\tr(\rho^2)$ and $\tr(\sigma^2)$, which are measures of the intra-cluster overlaps, and the term $\tr(\rho\sigma)$, which is the inter-cluster overlap. Looking at the upper-right panel, we see that the average positive pair clustering $\overbar{\tr(\rho^2)}$ increases rapidly at the start of training, before steadying at a value around 0.85. This demonstrates one mechanism by which $\bar{D}_{\text{HS}}$ increases, through the QNN producing representations which group positive pairs close together in Hilbert space. The bottom panels of Fig. \ref{fig:training} show the average negative pair clustering $\overbar{\tr(\sigma^2)}$ and average negative-positive pair overlaps $\overbar{\tr(\rho\sigma)}$, which decrease consistently throughout training. This demonstrates a second behaviour, whereby the QNN produces representations in which negative pairs are well separated. We note that these two values are very similar, which occurs in our self-supervised learning algorithm because of both the need to average over all positive pairs and because of the fixed size of $\rho_{i}$. Thus, in the limit $N\rightarrow \infty$ the ensemble $\sigma_i$ contains the entire batch and both metrics are effectively measuring the clustering of all data points. 

Overall, Fig. \ref{fig:training} shows that the quantum component of the encoder contributes to the overall learning process, despite the network's parameters being optimised explicitly in a classical space.
It is notable that the training time presented here is significantly less than classical benchmarks, which would typically be 100s of \textit{epochs}. Due to its technological infancy, executing quantum circuits on real or simulated hardware is computationally expensive. Thus, the 1-2 epochs of training used in this work represents the limit of our current experiment, although we expect this to improve dramatically in the coming years with the release of GPU-enhanced simulators~\cite{patti2021variational}. This also justifies our choice of dataset, since CIFAR-10 is both a modern relevant dataset~\cite{huang2019gpipe, cubuk2019autoaugment, phong2020rethinking} yet contains few enough images that we can complete at least one epoch.

\subsection{Linear probing}\label{subsec:linear_probing}

Once training is complete, we require a way to test the quality of the image representations learnt by the encoder. Specifically, a good encoding will produce representations whereby different classes are linearly separable in the representation space~\cite{kolesnikov2019revisiting}. Therefore, we numerically test the encoder using the established linear evaluation protocol~\cite{kolesnikov2019revisiting}, in which a linear classifier is trained on the output of the encoder network, whilst the encoder is frozen to stop it training any further. Once this linear probe experiment has trained for 100 epochs, we apply the whole network to unseen test data and record the classification accuracy.

\subsection{Quantum and classical results on the simulator}\label{subsec:main_results}

We repeat training, this time with the first five classes of CIFAR-10 and a projection head. We train models with three different types of representation networks; classical MLP with bias and Leaky ReLU activation functions after each layer, quantum trained on a statevector simulator and quantum trained on a sampling-based simulator. We choose the representation networks to be width $W=8$ in order to minimise the simulation overhead, whilst still being in a compression regime where training is stable (see Appendix \ref{app:classical_ablation}). Quantitatively, this means our two-layer classical and quantum representation networks have 144 and 32 learnable parameters respectively.

\begin{figure}
	\centering
	\includegraphics[width=\linewidth]{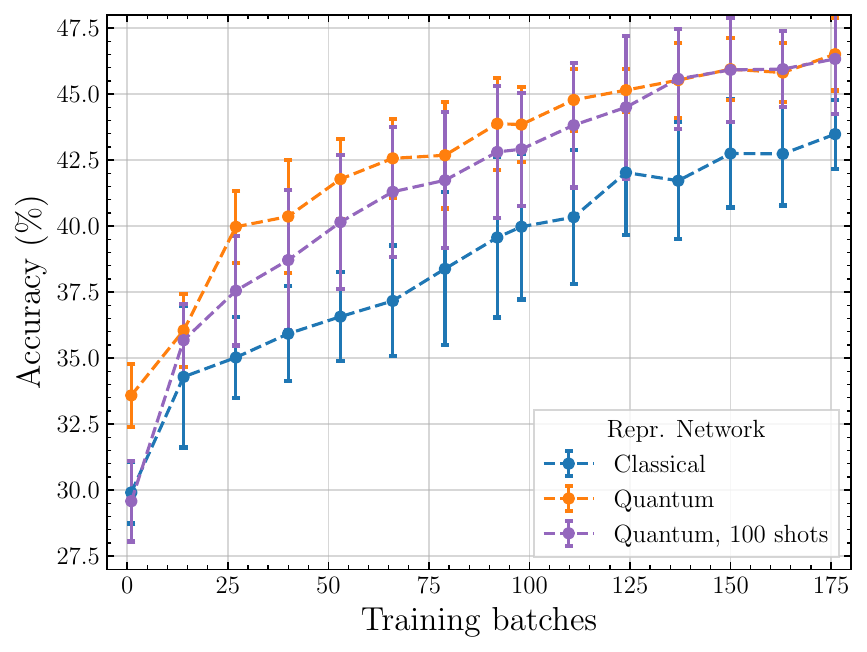}
	\caption{Classification accuracy achieved in linear probing experiments using the encoder at checkpoints across self-supervised training. Comparison between models trained with a classical representation network (blue), quantum representation network evaluated on a statevector simulator (orange) and quantum representation network evaluated on a sampling-based simulator with 100 shots (purple). The markers show the average of six independently trained models, whilst the error bars show one standard deviation.}
	\label{fig:main_result}
\end{figure}

Fig. \ref{fig:main_result} shows the result of linear probe experiments at checkpoints across 176 batches of contrastive training. We find that when the quantum circuits are evaluated using a statevector simulator, the quantum representation network produces higher average accuracy on the test set than the equivalent classical network at all points probed throughout training, and is separated by more than one standard deviation for over half of these. The highest accuracy is obtained at the end of training, where the quantum model achieves an accuracy of $(46.51\pm1.37)\%$ compared to $(43.49\pm1.31)\%$ for the classical model. In these results, the confidence interval corresponds to one standard deviation on the mean of six independently trained models. Furthermore, we also find that this numerical advantage holds for a range of smaller width models and is highly dependent on the correct choice of ansatz, more details of which are given in Appendices \ref{app:quantum_ablation} and \ref{app:ansatz_comparison}.

Subsequently, we explore whether using a finite number of shots limits this advantage. We train another quantum model on a simulator where the expectation values of measured qubits are sampled from 100 shots, both in the forward pass (generating the representations) as well as the backwards pass (calculating gradients). We find that beyond the first batch, the average accuracy of this model is still above what is achieved by the classical representation network, reaching $(46.34\pm2.07)\%$ by the end of training. Significantly, this matches the performance of the statevector simulator, which represents the limit of infinite shots, demonstrating resilience of our scheme to shot noise. However, we note that the additional uncertainty introduced by the sampling does manifest as a larger standard deviation between repeated runs, compromising the consistency of the advantage.

\subsection{Real device experiments}\label{subsec:real_device}

In section \ref{subsec:main_results}, we showed that a numerical advantage can be achieved for self-supervised learning with a quantum representation network, even when sampling the quantum circuits with only 100 shots. However, it does not follow that such an improvement can necessarily be realised on current quantum devices. The biggest barrier to this is the complex noise present on quantum hardware, a product of both the finite lifetime that qubits can be held in coherent states for and imperfections in the application of logic gates. To this end, we test the ability of real devices to accurately prepare representations produced by a pretrained quantum model and how this changes downstream accuracy on the test set.

We construct a linear probe experiment with a quantum representation network and load in weights from the best performing pretrained model in which circuits were evaluated with 100 shots. Freezing all of the layers so that the entire network no longer trains, we repeat classification of images from the test set, however this time the circuits are executed on IBM's 27-qubit \textit{ibmq\_paris} quantum computer. To reduce the number of gates, particularly SWAP operations caused by a mismatch between the ansatz and physical qubit connectivities, the circuits are recompiled using incremental structural learning~\cite{jaderberg2020minimum} before execution, the details of which can be found in Appendix \ref{app:ISL}.

\begin{figure}
	\centering
	\captionsetup[subfigure]{margin={1.7cm}}
	\subfloat[\label{fig:classical_matrix}]{
		\includegraphics[height=38mm,valign=b]{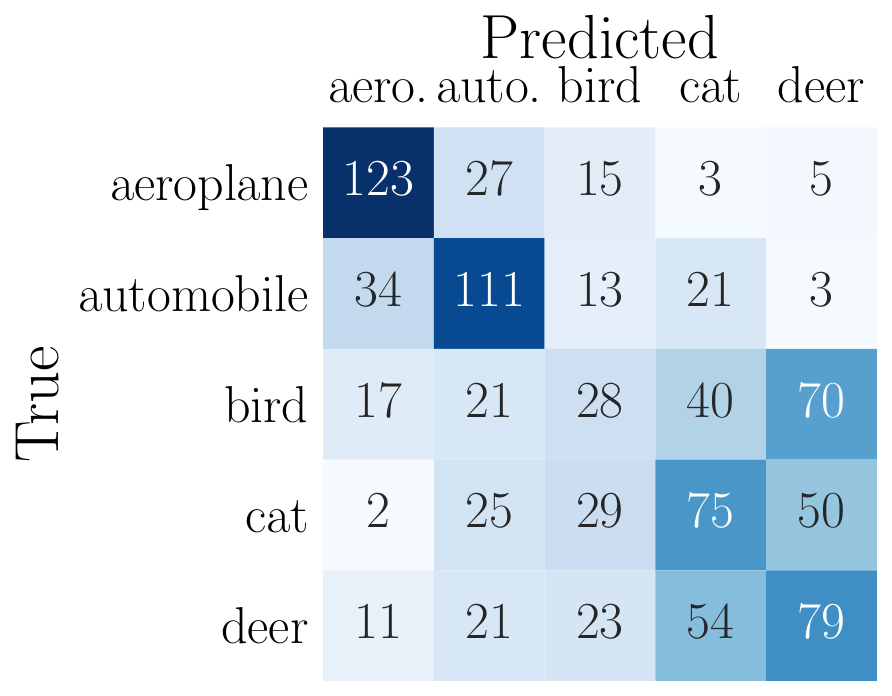}
	}
	\captionsetup[subfigure]{margin={0cm}}
	\subfloat[\label{fig:quantum_matrix}]{
	    \includegraphics[height=34.96mm, valign=b]{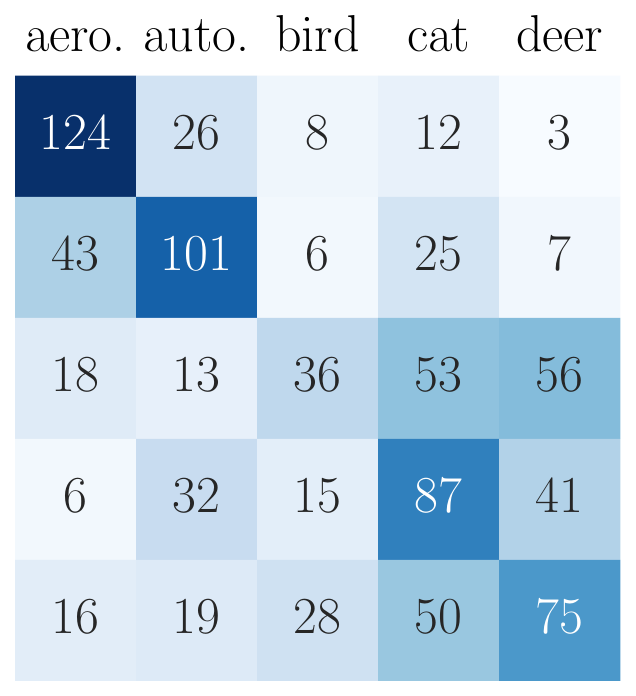}
	}	
	\caption{Confusion matrix from classifying 900 images using the best performing (a) classical model evaluated on a classical computer (b) quantum model evaluated on a real quantum computer with 100 shots per circuit.  For a given true label (rows) and predicted label (columns), the number in each box shows the total number of times that prediction was made.
		\label{fig:confusion}}
\end{figure}

Fig. \ref{fig:classical_matrix} shows the result of classifying 900 images randomly sampled from the test set, using the best performing classical model and evaluated on a classical computer. Fig. \ref{fig:quantum_matrix} shows the result when classifying the same images using the best performing 100-shot quantum model, evaluated on \textit{ibmq\_paris}. Overall, the classical and quantum models achieve an accuracy of $47.27\%$ and $47.00\%$ respectively. Excitingly, this demonstrates that in this experiment, error induced by noise on the quantum computer is able to be offset by the enhanced theoretical performance of quantum neural networks, provided the circuit depth is reduced with recompilation techniques. Furthermore, in both setups the most correctly predicted class was aeroplanes ($71.1\%$ and $71.7\%$) whilst the most incorrectly predicted class was birds ($15.9\%$ and $20.5\%$), both of which the quantum model performed better on. We propose that birds and deer were most likely to be mistaken with one another due to the images sharing a common background of the outdoor natural environment.

\section{Conclusion} \label{sec:conclusion}

In this work, we propose a hybrid quantum-classical architecture for self-supervised learning and demonstrate a numerical advantage in the learning of visual representations using small-scale QNNs. We train quantum and classical neural networks together, such that encodings are learnt that maximise the similarity of augmented views of the same image in the representation space, as well as implicitly in Hilbert space. After training is complete, we determine the quality of the embedding by tasking a linear probe to classify images from different classes. We find that an encoder with a QNN acting in the representation space achieves higher average test set accuracy than one in which the QNN is replaced by a classical neural network with equivalent width and depth, even when evaluating quantum circuits with only 100 shots. We note that although making such a comparison has been established in previous works~\cite{abbas2021power}, how to fairly compare quantum and classical neural networks still remains a significant open question.

We then apply our best performing pretrained classical and quantum models to downstream classification, whereby the quantum circuits were evaluated on a real quantum computer. The observation of a quantum predictive signal with equivalent accuracy to that of the classical model, despite the complex noise present on current quantum devices, is representative of the potential practical benefit of our setup. If recent progress in superconducting qubit hardware continues~\cite{arute2019quantum,kjaergaard2020superconducting,jurcevic2021demonstration}, it is possible that QNNs running on real devices will outperform equally sized classical neural networks in the near future in this experiment.

One advantage of the hybrid approach taken in this paper is the resulting flexibility in how much of the encoder is quantum or classical. In fact, there now exist numerous software solutions for producing and testing such hybrid architectures~\cite{Qiskit,bergholm2018pennylane,broughton2020tensorflow}. As the quality and size of quantum hardware improves, our scheme allows classical capacity to be substituted for quantum, eventually replacing ResNet entirely. By optimising directly for the Hilbert-Schmidt distance, it is also possible with a fully quantum encoder to apply our setup to problems in which the data is itself quantum~\cite{sentis2012quantum, alvarez2017supervised, amin2018quantum,gong2022quantum, szoldra2022unsupervised}. Promisingly, in this regime it may prove that the advantage observed in this work is further extended, given the ability of a quantum model to inherently exploit the dimensionality of the input~\cite{sentis2019unsupervised}. Recently developed data sets consisting of entangled quantum states~\cite{perrier2021qdataset,schatzki2021entangled} serve as an obvious target for such work. In this case contrastive augmentations could be quantum operations that change the state but conserve the properties of interest, for example LOCC operations that do not affect the amount of entanglement in the system. With classical contrastive learning having been applied to non-visual problems in biology~\cite{lu2020self} and chemistry~\cite{jaeger2018mol2vec}, our work provides a strong foundation for applying quantum self-supervised learning to fundamentally quantum problems in the natural sciences~\cite{cong2019quantum}.

An open question remains as to whether a general quantum advantage for self-supervised learning may prove possible~\cite{arute2019quantum,zhong2020quantum}, in which no classical computer of any size can produce accuracies equal to that of a quantum model. In~\cite{abbas2021power, abbas2021effective}, the authors define \textit{effective-dimension}, a metric measuring the expressive power of classical and quantum neural networks. In general, quantum models are able to achieve a higher effective dimension, and therefore capture a larger space of functions, than classical models with comparable width and number of parameters. Although it does not necessarily increase monotonically, the effective dimension of quantum models can remain larger than classical as the model and data set size are increased. Such behaviour indicates that the expressive power available to QNNs may allow for an advantage over classical neural networks, particularly for a problem such as self-supervised learning where highly expressive, large capacity models are believed to be particularly important for achieving highly accurate predictions~\cite{chen2020simple}.

Achieving experimental quantum advantage would require, as a minimum, a QNN with width greater than 60 qubits, such that the dimensionality of the accessible feature space becomes classically intractable. Furthermore, the QNNs would need to be trained on real devices, which remains a challenge due short qubit lifetimes and low gate fidelity. Therefore, considerable research still remains into the scalabiltiy of our scheme, which was only demonstrated at the small sizes feasible on current quantum hardware. Promisingly however, our method can be adapted to use different QNN structures that avoid the scaling issue of barren plateaus~\cite{pesah2021absence, grant2019initialization, cerezo2021cost}, which could be tested already using more efficient simulators~\cite{luo2021quantum}. Looking forward, the rate at which quantum hardware continues to progress provides the possibility of representing intractable distributions using QNNs. In this way, quantum computers may yet push self-supervised learning beyond the performance afforded by classical hardware.

\section{Data availability}
Data used to generate the above figures are available upon request from the authors.

\section{Code availability}
The code used to train the models described in this work can be found at \url{https://github.com/bjader/QSSL}. The code used to incorporate and train Qiskit quantum neural networks into PyTorch can be found at \url{https://github.com/bjader/quantum-neural-network} and is required to build quantum representation networks.

\begin{acknowledgments}
	B.J., L.W.A., M.K. and D.J. acknowledge support from the EPSRC National Quantum Technology Hub in Networked Quantum Information Technology (EP/M013243/1) and the EPSRC Hub in Quantum Computing and Simulation (EP/T001062/1). M.K. and D.J. acknowledge financial support from the National Research Foundation, Prime Ministers Office, Singapore, and the Ministry of Education, Singapore, under the Research Centres of Excellence program. W.X. and S.A. are supported by EPSRC grant Seebibyte (EP/M013774/1) and Visual AI (EP/T028572/1).
\end{acknowledgments}

\appendix

\section{Contrastive Loss Function}\label{app:contrasitve_loss}

Here we formally define the process of contrastive learning. Let us have an augmentation function $\xi(\cdot; a)$. This augmentation combines cropping, rotation, Gaussian blurring and colour distortion of the image, and the amount by which each of these operations is performed is governed by a list of continuous random variables $a$. Each time we apply an augmentation, we randomly sample $a$ from a distribution $A$ such that applications of the augmentation function are independent from one another.

For a particular image $\vec{x}_i$, we now have a pair of views $\mathcal{P}_i = \set{\xi(\vec{x}_i; a_1), \xi(\vec{x}_i;a_2) \mid a_1, a_2 \sim A}$ which came from the same base image. We call this a positive pair. We define the negative pairs as the set of all augmented versions of different images.

During contrastive training, all augmented views within the batch are passed through our architecture. The encoder network $f(\cdot): \vec{x} \rightarrow \vec{y}$ and projection head $g(\cdot): \vec{y} \rightarrow \vec{z}$, are applied to give outputs $\vec{z}_i^\alpha = g(f(\xi(\vec{x}_i; a_\alpha))); \: a_\alpha \sim A$ for each of the two arms (labelled by $\alpha=1,2$).

For simplicity, we define the NT-Xent loss for each input data separately labelled by index $i$ as follows. The overall loss function corresponds to the sum of these terms over all input images (and correspondingly defined positive and negative pairs). A single term in the loss term is given by

\begin{equation}
    \mathcal{L}_i = \log \dfrac{-\exp\left(\vec{z}^1_i\cdot \vec{z}^2_i/\tau\right)}{\sum
    _{\substack{j,k \in \{1,\dots,N\} \\ \alpha,\beta \in \{1,2\}}}
    \exp\left(\vec{z}^\alpha_j\cdot \vec{z}^\alpha_k/\tau\right)},
\end{equation}

where $i=1,2,\dots,N$ labels the input image and $\alpha,\beta=1,2$ labels the (arbitrary) distinction between the first and second augmentation making up the positive pair. The overall loss $\mathcal{L}$ is given by the sum over each of $i$.

\section{Training hyperparameters} \label{app:hyperparameters}

Throughout this work, the training parameters used are; batch size: 256, optimiser: ADAM~\cite{kingma2014adam} with $\beta_{1}=0.9, \beta_{2}=0.999$, learning rate: $10^{-3}$, weight decay: $10^{-6}$ and softmax temperature: 0.07.

\section{Classical width ablation}\label{app:classical_ablation}

In order to incorporate QNNs that can be run on current quantum devices into contrastive learning, a compression of the feature vector is required after ConvNet. Since this would not be necessary in a purely classical setting, its impact on final performance is not well understood. To this end, we perform a study of the accuracy achieved by models with different representation network widths. We do this with classical representation networks to remove the quantum specific considerations of statistical noise and optimal circuit architecture, focusing purely on width. The classical representation network is a two-layer, width $W$ MLP, with Leaky ReLu activation functions after each layer and with bias.

\begin{figure}
	\centering
	\includegraphics[width=\linewidth]{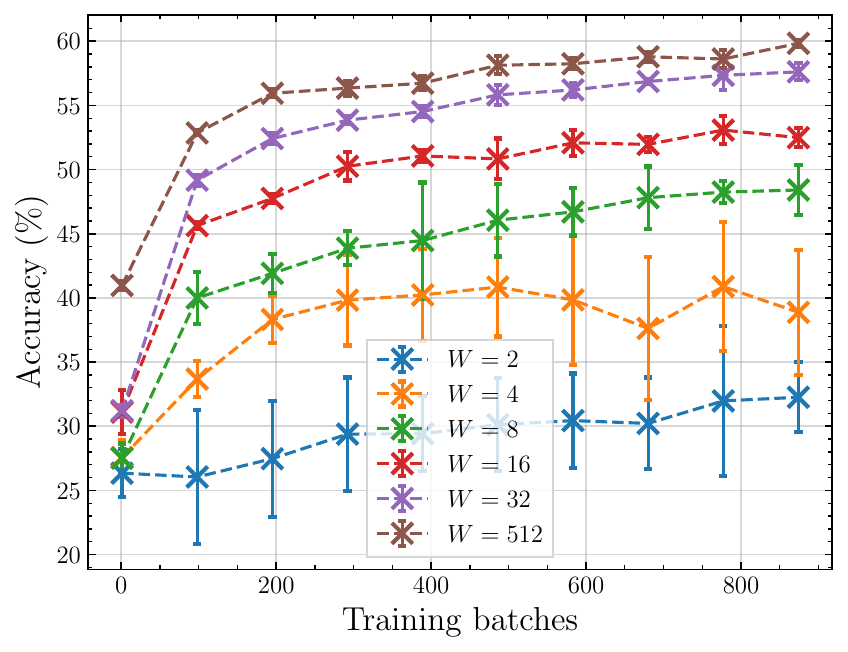}	\caption{Classification accuracy achieved in linear probing experiments by classical representation networks with varying network widths at checkpoints across self-supervised training. The markers show the average of three independently trained self-supervised models and linear probe experiments, whilst the error bars show one standard deviation.}
	\label{fig:classical_width_ablation}
\end{figure}

Each model is trained on the first five classes of the CIFAR-10 dataset and a linear probe experiment evaluates the performance at regular checkpoints during training. Fig. \ref{fig:classical_width_ablation} shows the result comparing models with different representation network widths, including the $W=512$ case which corresponds to no compression. Starting from $W=2$, we see that increasing the width of the representation network improves the test accuracy. Furthermore, we find that $W=8$ is the lowest width network in which test accuracy retains the same qualitative behaviour as the uncompressed network. Therefore, in our proof-of-principle quantum experiments, we use an eight width representation network corresponding to eight qubits.

\section{Quantum and classical results at different widths}\label{app:quantum_ablation}

In section \ref{subsec:main_results} we demonstrate that for an architecture with a $W=8$ representation network, using a QNN to form a hybrid model leads to higher performance in linear probing experiments than the purely classical case. Here we supplement this with additional experiments for the $W=2$, $4$ and $6$ cases alongside an equally sized classical comparison for each one. The same problem setup and training parameters are used as in Fig.~\ref{fig:main_result}.

\begin{figure}
	\centering
	\includegraphics[width=\linewidth]{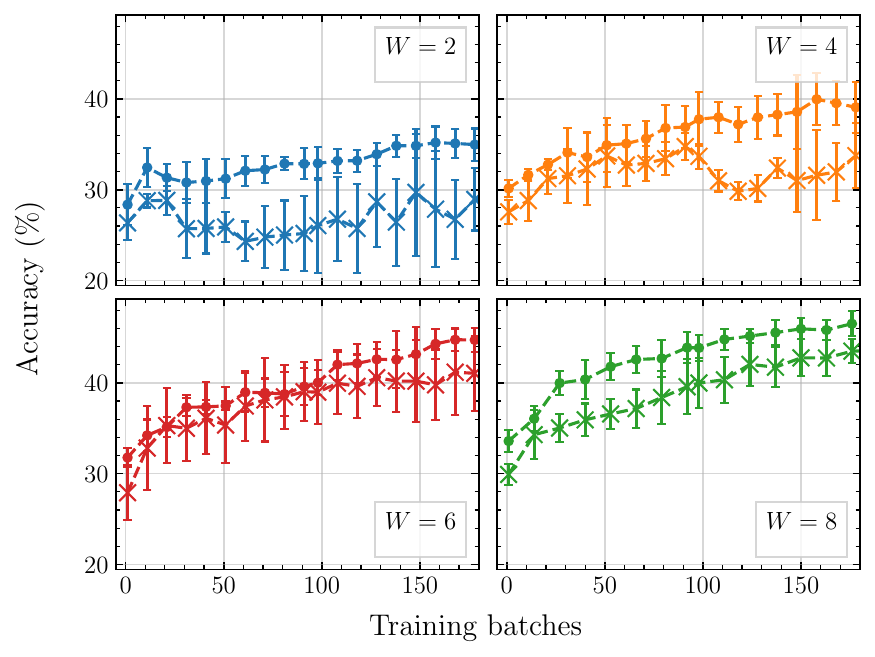}
	\caption{Classification accuracy achieved in linear probing experiments for quantum and classical models with representation network widths 2 (upper left), 4 (upper right), 6 (lower left) and 8 (lower right). Within each graph, the circle (crossed) markers represent models trained with quantum (classical) representation networks. The markers show the average accuracy of three independently trained models, with error bars of one standard deviation, except for the $W=8$ case which displays the same data as Fig. \ref{fig:main_result}.}
	\label{fig:quantum_classical_ablation}
\end{figure}

Fig.~\ref{fig:quantum_classical_ablation} shows the accuracy achieved by these additional models in linear probing experiments at intervals across training, as well as the $W=8$ results from the main text. Focusing on the circle markers representing the quantum models, we see that the accuracy improves consistently when increasing the QNN width. This matches the behaviour of the classical models, represented in this figure by the crossed markers, illustrating that our intuition for how compression of the network affects performance can be applied to both the quantum and classical regimes. Secondly, we compare between quantum and classical models of the same width, as shown by the lines of the same colour. Here we see that for the new cases of $W=2,4,6$, there is a numerical improvement in the average accuracy achieved across all training checkpoints sampled, consistent with the $W=8$ case. Whilst these are still small models, they provide further impetus to consider whether this improvement would remain for models with width $W>8$, eventually competing directly with the uncompressed SimCLR algorithm at $W=512$. Looking forward, testing this hypothesis towards the $W=60$ qubit range may be possible with more efficient simulators~\cite{luo2020yao, suzuki2021qulacs} as well as by employing training shortcuts such as calculating gradients directly with the quantum state rather than using the parameter shift rule~\cite{bausch2020recurrent}.

\section{Performance of alternative ansatz}\label{app:ansatz_comparison}

\begin{figure}
	\centering
	\vspace{0.25cm}
    \includegraphics[width=\linewidth]{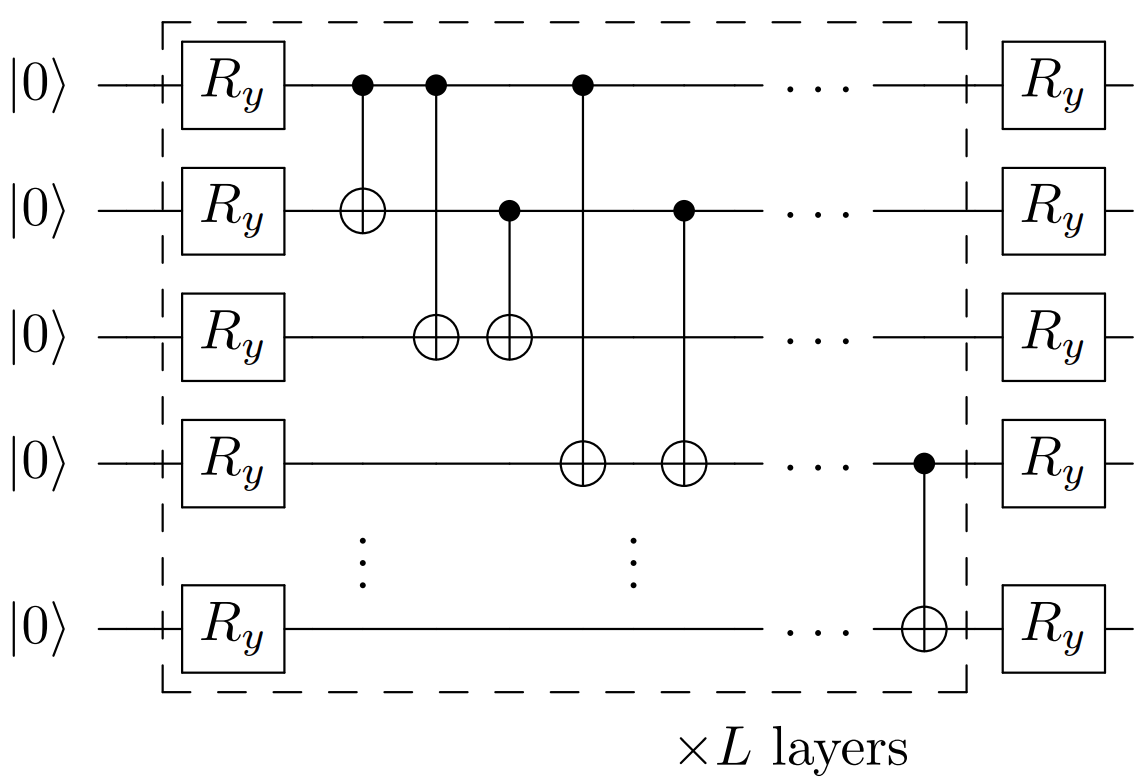}
    	\caption{Alternative variational ansatz. Each layer consists of a single qubit $\hat{R}_y$ rotation on each qubit, followed by CNOT gates connecting all qubits to each other. After all layers have been applied, a final set of $\hat{R}_y$ rotations are applied. Every rotation gate is parameterised by a different variational parameter.}
    \label{fig:abbas}
\end{figure}

In section \ref{sec:results} all QNNs are constructed using the variational ansatz seen in Fig. \ref{fig:sim_circ_14}, which connects the qubits in a ring of parameterised controlled rotation gates. Here we introduce a second ansatz, as seen in Fig. \ref{fig:abbas}, which is different in that it connects all of the qubits together and only has single qubit parameterised gates. Notably, this ansatz was recently shown to exhibit a larger effective dimension when applied to supervised learning than equivalent classical networks~\cite{abbas2021power}. Therefore, we test whether this circuit structure is also a good candidate for improved performance in a self-supervised setting.

\begin{figure}
	\centering
	\includegraphics[width=\linewidth]{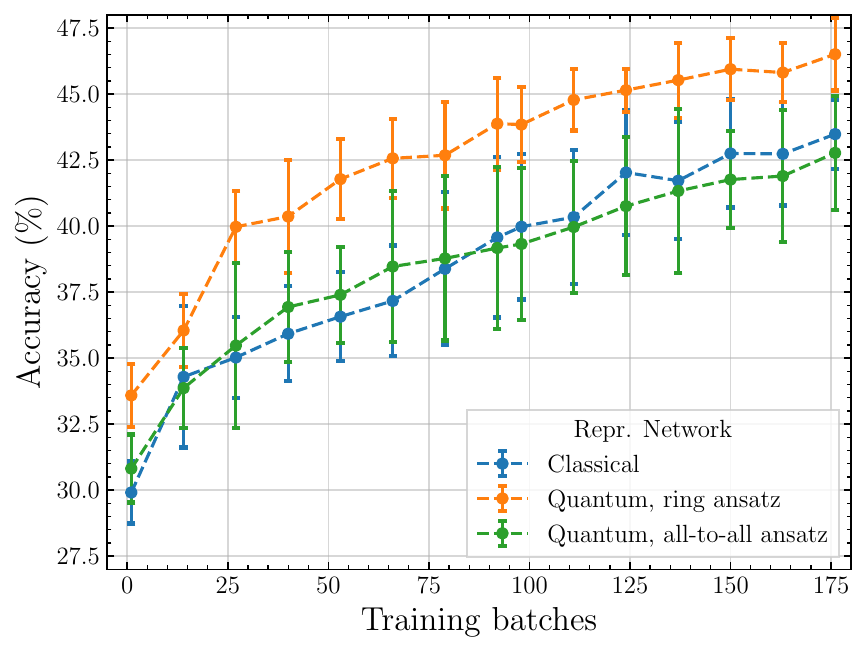}
	\caption{Classification accuracy achieved in linear probing experiments using the encoder at checkpoints across self-supervised training. Comparison between models trained with a classical representation network (blue), quantum representation network with the ring ansatz (orange) and quantum representation network with the all-to-all ansatz (green). All quantum circuits were evaluated on a statevector simulator. The markers show the average of six independently trained models, whilst the error bars show one standard deviation.}
	\label{fig:ansatz_comparison}
\end{figure}

We train a model with a quantum representation network structured as the new all-to-all ansatz, simulated on a statevector simulator. The dataset consists of the first five classes of CIFAR-10 and the model is trained with a projection head. Importantly, for a fair comparison, we apply three layers of the all-to-all ansatz, so that it has the same number of learnable parameters as two layers of the ring ansatz. The result of the linear probe experiments can be seen in Fig. \ref{fig:ansatz_comparison}, along with the previous models for comparison. We see that for the all-to-all ansatz, test accuracy is no higher than the classical model beyond the statistical variance of repeating training with different initial parameters, and below the ring ansatz. Indeed, by the end of training, the all-to-all ansatz achieves a final accuracy of $(43.46\pm1.68)\%$, which is similar to the classical model. Thus, we show that achieving an advantage using quantum neural networks in contrastive learning is highly dependent on the correct choice of quantum circuit structure.

\section{Recompilation of quantum neural networks}\label{app:ISL}

When executing QNNs on the \textit{ibmq\_paris} device, translating the ring topology of our variational ansatz to the honeycomb structure that the qubits are physically connected by requires a significant number of SWAP operations. Quantitatively this increases the number of two-qubit gates in the circuit from 16 to 143, which poses a significant challenge to obtaining a predictive signal beyond random noise since the total circuit error scales exponentially with the number of gates. To mitigate this, for each image evaluated we approximately recompile the QNN using incremental structural learning (ISL)~\cite{jaderberg2020minimum}, adapted so that only two-qubit connections available on the real device can be applied. Using this method, for over half of the executed circuits, an equivalent circuit is found which produces the same statevector with at least $99\%$ overlap using on average 14 CNOT gates. For the remaining images, we apply ISL once again, but this time without any constraints on the connectivity of the circuit. This produces a shallower equivalent circuit with at least $99\%$ overlap using on average 8 CNOT gates. Although some of these two qubit gates require SWAPs when implemented on the real device, they still represent a significant reduction in the depth of the circuit and total error incurred.

\bibliography{bibliography.bib}

\end{document}